\documentclass[12]{article}
\usepackage{graphics}

\begin{document}
\bibliographystyle{h-physrev}
\input{epsf}

\title{Superconductivity in a Dilute Two Dimensional Electron System}
\author{A.Agarwal\thanks{abhishek@pas.rochester.edu} \\
\and
 S.G.Rajeev\thanks{rajeev@pas.rochester.edu} \\
University of Rochester. Dept of Physics and Astronomy. \\
Rochester. NY - 14627}
\maketitle

\begin{abstract}
We develop a finite temperature mean field theory in the path integral
picture for an extremely dilute 
system of
interacting Fermions
in a plane.  In the limit of short ranged interaction, the system is
shown to undergo a phase transition to a
superconducting regime. Unlike the well known BCS transition for
metals, this phase transition is found to be very
sensitive to the two dimensional nature of the problem. A quantitative
estimate of this sensitivity is carried out by repeating the analysis at
non - zero thickness. The validity of the mean field results are also
proved using renormalization group techniques.
\end{abstract}

\section{Introduction}
It is well known from the theory of superconductivity\cite{Abrikosovetal},
that the low
energy effective theory for electrons in metals is susceptible to a BCS
phase transition at $T = 0$. The low energy theory
which involves only the modes close to the Fermi surface ($K_F$) turns out
(in any dimension $D \geq 2$), to be an effective one dimensional theory
that describes  a large number (N) species of Fermions satisfying a linear
energy momentum dispersion relation. The inter - electron interaction for
this theory
turns out to be short ranged and attractive\cite{Shankarbcs,
Polchinski:bcs}. The one dimensional nature 
of the effective theory and the linear energy momentum dispersion
relation, are key ingredients that make the theory asymptotically
free,  while the attractive nature of the
inter - electron potential is responsible for the formation of the gap in 
the energy spectrum. The effective one dimensional
nature of the BCS theory is related to the fact that in considering modes
close to the Fermi sea, (a natural cut off for the theory), one finds
that all the directions in momentum space tangential to $K_F$ are
degenerate, in the sense that the energy depends only on the directions
normal to the Fermi sea. Hence one naturally obtains a large degeneracy,
denoted by the number N, which turns out to be the area of the Fermi
surface measured in units of the cut off. Moreover, this dimensional
reduction to one dimension can be carried out in a manner which is
insensitive to the true dimensionality of the problem which is one of the
reasons behind the universality of the BCS transition\cite{Shankarbcs,
Abrikosovetal, Feldmanetal}.

The attractive inter - electron potential is also generic to all metallic
systems in $D \geq 2 $. This can be best understood by appealing to
renormalization group (RG) arguments. RG analysis shows that due to
coulomb screening all long ranged interactions (repulsive or otherwise) 
behave in a 
screened or short ranged manner. If one now decomposes the short ranged
interaction into its various angular momentum components, then the RG
flows for the various components couple to each other. The structure of
these coupled equations suggests that it is always possible to find some
value of $l$ (the angular momentum ), for which the effective coupling for
that sector will flow to negative values. The couplings that do run
negative , i.e become attractive, lead to a BCS transition, while the
others renormalize to zero. This is the so called Kohn - Luttinger
effect\cite{Luttinger1, Luttinger2, Baranovetal}.
The Kohn - Luttinger effect, and the dimensional reduction to one
dimension make the BCS phenomenon truly universal, irrespective of the
true physical dimensionality of the metallic systems or the microscopic
details of the inter electron interaction.

Recently there has been much theoretical work towards  understanding
the behavior of two dimensional dilute electron - hole
systems\cite{Abrahamsetal:metal2d, Phillips1}.  Whether or not such
systems can undergo a phase transition
to a conducting or superconducting regime, and the dependence on the true
physical
dimensions of such possible phase transitions  is the object of much
speculation, as these systems arise naturally in the context of high $T_c$
superconductors. The physical simplifications mentioned before in the
context of the usual BCS theory do not apply to these dilute systems,
where the physical momentum scale $\Lambda >> K_F$. For such systems it does not make sense to
linearize the energy - momentum dispersion relation  around  $K_F$ and
reduce 
to a one dimensional
problem. We present
here the analytical study of such a dilute system of electrons,
interacting through an attractive short ranged potential. We carry out the
study at finite temperature in the absence of impurities or disorder, and
find that such systems undergo a phase transition to a superconducting
state for arbitrarily small values of the attractive coupling. Moreover we
find that they do so in a manner which is extremely sensitive to the two
dimensionality of the problem. 

We will  start with a short review of the quantum mechanical two body
problem
involving an attractive delta function potential in two dimensions, as it
is an
excellent toy model that encapsulates all the essential features of the
corresponding many body problem, eg. the formation of bound states and
asymptotic freedom. We shall then focus on the second quantized version of
the problem, which will be  analyzed in the path integral formalism. This will be followed
by an exploration of the sensitivity of the BCS like phase transition to
two dimensionality. In a separate section we shall justify the use of mean
field theory in the context of this problem using renormalization group
techniques.

\section{Short Ranged Attraction in two Space Dimensions and
Asymptotic Freedom}

{\bf The Two body problem:}\\
Let us consider a system of two particles interacting through an
attractive short ranged potential in two dimensions\cite{Huangquark,
Rajeevgupta, Rajeevhenderson}. This two body  
system has an ultraviolet divergence, and the subtleties
associated with the renormalization of theories  which have one marginal
parameter. The Schroedinger equation for this system (in momentum space) 
is:
\begin{equation}
(\frac{p^2 }{2m} - E)\Psi _{\Lambda }(p) -g(\Lambda )\rho _{\Lambda }(p)
\int \rho _{\Lambda }(q)\Psi _{\Lambda }(q)[d^2q] = 0
\end{equation}
Note: $[d^2q]$ is a short hand for $\frac{d^2q}{2\pi ^2}$.\\
The equation above is regularized by the presence of $\rho _{\Lambda }(p)
=
\Theta (|p|<\Lambda )$. In the absence of the regularisation, the
equation is ill defined, because the ground state energy of
the system is not bounded below. This is an artifact of the scale
invariance of the equation which implies that if $E$
is an eigenvalue,
then so 
is $sE$, where $s$ is a scaling parameter. Hence the bound state energy
can 
be either $0$, or $-\infty $. We can make sense
of this scenario, by formulating the basic dynamical equation with a
cut-off, as above, and ask the question as to how  the dimensionless
coupling constant $g$ depend on $\Lambda $, such that all the physical
quantities have a finite limit as $\Lambda \rightarrow \infty $.   
Solving for $\Psi _{\Lambda }(p)$ gives:
\begin{equation}
\Psi _{\Lambda }(p) = \frac{A_{\Lambda }}{\frac{p^2}{2m} - E}\rho
_{\Lambda }(p)  
\end{equation}

Where $A_{\Lambda } = g(\Lambda )\int \rho _{\Lambda }(p)\Psi _{\Lambda
}(p)[d^2p]$. Putting the expression for $\Psi _{\Lambda }(p)$ in the
equation for $A_{\Lambda }$, we have:
\begin{equation}
g^{-1}(\Lambda ) - \int \frac{\rho ^2_\Lambda (p)}{\frac{p^2}{2m} -
E}[d^2p]
= 0
\end{equation}  
This expression for $g^{-1}(\Lambda )$ is log-divergent. The situation can
be remedied if we choose $g(\Lambda )$ in the following way:
\begin{equation}
g^{-1}(\Lambda ) = \int \rho ^2 _{\Lambda }(p)\frac{1}{\frac{p^2}{2m} +
\frac{\mu ^2 }{2m}}[d^2p]
\end{equation}
With this choice of $g$, Eqn$(3)$ above has a finite limit as $\Lambda
\rightarrow \infty $, and the resulting theory is asymptotically free.   
The essence of renormalization here is the trading of the bare coupling
constant $g$ in favor of the number $\mu $, which has the dimensions of
momentum. This number is the true physical parameter of the theory, which 
describes the strength of the interaction between the two particles.

The nature of the divergence, does not change if we make the transition
from the two body problem, to the corresponding many body problem of
non-relativistic particles in a plane. Indeed, it can be
shown, that the divergence present at the level of the mean field theory
of the many body problem is the only divergence that requires
renormalization. In other words, renormalizing the mean field theory is
enough to cast the problem in a finite form. Moreover renormalization
group arguments (which we present in the last section)show that the
predictions of the mean filed analysis for this problem are qualitatively
correct. In the next section  we will  generalize the two body
Hamiltonian
given above to the corresponding many body case, and recast the problem of
evaluating the thermodynamic partition function for this Hamiltonian in the
language of path integrals. In the next section we shall use this path
integral point of view to construct the mean field theory for the problem.

{\bf The many body problem:}\\
The generalization of the Hamiltonian
given in Equation (1) to a second quantized language is,
\begin{equation}
H = \int [d^Dr]\Psi ^{\dagger a_1 }(r)[-\nabla ^2]\Psi _{a_1
}(r)
 -\alpha(\Lambda )\int[d^Dr]\Psi ^{\dagger
a_1}(r)\Psi ^{\dagger
a_2}(r)\Psi _{a_2}(r)\Psi _{a_1}(r)
\end{equation}
We are keeping the dimensionality of space ($D$) arbitrary for the moment.
This Hamiltonian can be written in a manner more suited to a mean field
analysis, by using  the following two
component spinors,

\begin{equation}
\Psi = \left( \begin{array}{c}\psi _{\uparrow} (r) \\ \psi
_{\downarrow}^{\dagger }(r) \end{array}\right)
\end{equation}
In terms of $\Psi$, the  Hamiltonian can be written as,
\begin{equation}
H = \int
d^D(r)\left[\Psi^{\dagger }( - \nabla ^2)\sigma _3 \Psi - \alpha \Psi
^{\dagger }\left( \begin{array}{cc}0 & \psi _{\downarrow}\psi _{\uparrow} 
\\ \psi ^{\dagger
}_{\uparrow}\psi ^{\dagger }_{\downarrow} & 0
\end{array} \right) \Psi \right]
\end{equation}
The composite operator  $\psi ^{\dagger } _{\uparrow}(r)\psi ^{\dagger
}_{\downarrow}(r)$, 
 and
it's
Hermitian conjugate appearing in the Hamiltonian above are the creation
and annihilation operators for the quasi particles of the theory, which
are scalars in the present case. The interesting physical features of the
theory, such as the formation of a gap in the spectrum are related to the
fact that $\langle \psi ^{\dagger }_{\uparrow} \psi ^{\dagger
}_{\downarrow} \rangle\neq o$ in the ground state.
Hence we would like to reformulate the problem in a language where these
composite operators are the chief dynamical variables. We shall do it
below in the finite temperature limit and in the path integral picture.
  
The finite temperature features of the theory can be probed by calculating
the thermodynamic partition
function $Tr[e^{-\beta H}]$ for the Hamiltonian H given above in (5). Now
for any
normal ordered Fermionic Hamiltonian one can reduce the problem of
calculating the partition function to the evaluation of a Path integral
over Grassmann variables, using the prescription,
\begin{equation}
Z = Tr[e^{-\beta H}]=\int e^{-\beta \int _0 ^1dt\int d^Dr[\Psi^{\dagger
}(rt)\frac{1}{\beta }\partial _t\Psi (rt) - H(\Psi ^{\dagger
}(rt),\Psi (rt))]}D[\Psi
^{\dagger }]D[\Psi ]
\end{equation}  
Here it is understood, that the arguments of H appearing in the path
integral are the Grassmann variables corresponding to the Fermion creation
and annihilation operators appearing in Equation (5). This functional
integral can be rewritten in the following way.
\begin{equation}
Z = \int e^{-\beta \int _0 ^1 dt\int d^{D} r(\Psi
^{\dagger
}(\frac{1}{\beta }\partial _t - \nabla ^2 )\sigma _3 \Psi + g \Psi
^{\dagger }\left( \begin{array}{cc} 0 & \phi \\ \phi ^{\dagger } & 0
\end{array} \right)\Psi + \frac{1}{2} |\phi |^2)}D[\Psi ^{\dagger }]D[\Psi
]D[\phi ^{\dagger }]D[\phi]
\end{equation}
The functional appearing in the exponential is to be thought of as the
action $S$ of an interacting theory.
Here $\phi $ is an auxiliary complex scalar field.
To see that this is the correct action, we can integrate out 
the scalar
fields, and recover an effective action for the Fermi
fields 
which is  the
action appearing in (8). i.e.
\begin{equation}
\int D[\phi ]D[\phi ^{\dagger }]e^{-S } = e^{-S_{Fermi}}
\end{equation}

where
\begin{equation}
S_{Fermi} = \beta \int_0 ^1 dt\int d^{D}r[\Psi ^{\dagger}(r)(\frac
{1}{\beta }\partial _t -\nabla ^2
)\sigma
_3
\Psi (r)- 2g^2\psi^{a \dagger}(r)\psi^{b
 \dagger }(r)\psi_{b}(r)\psi_{b }(r)
\end{equation}
It is now obvious that $S_{Fermi }$ is the action generated by the
Hamiltonian in (5), when translated to the path integral picture using
(8) if we identify $2g^2 $ with the coupling constant $\alpha(\Lambda )$
appearing in (5).  A comparison between equations (7) and (9) makes it
obvious that $\phi $ describes the quasi bosons of the theory,  i.e $\phi
=\psi _{\downarrow} \psi _{\uparrow} $. Reformulation of the theory in
terms of the composite
operator describing the quasi scalars can now be accomplished, by
integrating out the Fermi fields appearing in (9). This produces the
effective action for the complex scalar field.
\begin{equation}
S_{Eff}[\phi, \phi^{\dagger }, \beta ] = \beta \frac{1}{2}\int d^Dr|\phi
|^2 -Tr\ln\left[ 1 +
\frac{g}{(\frac{1}{\beta }\partial _t - \nabla ^2)\sigma _3
}\left( \begin{array}{cc} 0 & \phi \\ \phi ^{\dagger } & 0 \end{array}
\right) \right]
\end{equation}
Transforming to Fourier space, we have;
\begin{equation}
Tr\ln(1 + \frac{g}{(\frac{1}{\beta }\partial _t - \nabla ^2 )\sigma _3 }
\left( \begin{array}{cc} 0 & \phi \\ \phi ^{\dagger } & 0
\end{array}\right) ) = \int [d^Dk]\ln \Pi _n(1 + \frac{\beta ^2g^2|\phi   
|^2}{(2n +   
1)^2\pi ^2 + \beta ^2k^4})
\end{equation}
The product on the left hand side is over the (odd) Matsubara frequencies.
The evaluation of this trace can be carried out as follows; let
\begin{equation}
A = \ln \Pi _n ( 1 + \frac{x^2}{(2n + 1 )^2\pi ^2 + y^2 })
\end{equation}
Where $x = \beta g |\phi | = \beta \Delta $, and $y = \beta k^2$. Now;
\begin{equation}
\frac{d}{dx^2}A = \Sigma _{-\infty }^{\infty } \frac{1}{(2n +1)^2\pi ^2
+ \beta ^2E^2} = 2 \Sigma _1 ^{\infty } \frac{1}{(2n -1 )^2 \pi ^2 + \beta
^2E^2}
\end{equation}
In the equation above, $E = (k^4 + \Delta ^2 )^{1/2}$.
Now recalling the identity
\begin{equation}
\tanh(\frac{\pi x}{2}) = \frac{4x}{\pi}\Sigma _1
^{\infty}\frac{1}{(2n-1)^2
+ x^2}
\end{equation}  
we get
\begin{equation}
\partial _{x^2} A = \frac{1}{2\beta E }\tanh(\frac{\beta E}{2})
\end{equation}
\begin{equation}
A = \frac{\beta }{2}\int _0 ^{g^2 |\phi |^2} \frac{\tanh(\frac{\beta
E(x)}{2})}{E(x)}dx^2
\end{equation}
Where $E(x) = (\frac{x^2}{\beta ^2} + \omega ^2(k))^{1/2}, \omega (k) = k
^2$.\\

{\bf Mean field theory:} Mean field theory corresponds to the
saddle point of the non - linear scalar field theory described by (12). To
get the saddle point equations, it is sufficient to consider a constant
value for the scalar field $\phi _c$. Using the form given in (18) for the
functional determinant, the effective action for a constant value of the
scalar field  is,

\begin{equation}
S_{Eff}[\phi _c] = \beta (\frac{|\phi _c |^2}{2} - \frac{1}{2} \int
\int_0^{g^2|\phi _c |^2}\frac{\tanh(\frac{\beta E }{2})}{E}[d^Dk]dx^2) = F
\end{equation}  
Where F denotes the free energy of the system.\\
The mean field (saddle point equations):\\ 
The equation for the saddle point is the equation for the extremum of the
effective action, or the Free energy, hence we have from the form of
$S_{Eff}$ above, 
\begin{equation}
\frac{\delta F}{\delta \phi _c } = 0 \Rightarrow 1 = g^2 \int
d^Dk\frac{\tanh (\frac{\beta (\Delta ^2 + \omega
^2(k))^{1/2}}{2})}{(\Delta
^2
+ \omega ^2(k))^{1/2}}
\end{equation}
This is the familiar gap equation, and $\Delta = |g|\phi _c $ has the
physical interpretation of the gap in the energy spectrum.
This equation is
divergent in any dimension  $D \geq 2 $. The divergence can be made
explicit
by considering the zero temperature $(\beta \rightarrow \infty )$ limit,
where it becomes,
\begin{equation}
g^2 \int [d^D]k\frac{1}{(\Delta ^2 +~ \omega ^2 (k) )^{1/2}} = 1
\end{equation}

In the case
of $D= 2$, which is the critical dimension for the theory,this ultraviolet
divergence is logarithmic, and it can be renormalized if we let
$g^{-2} \sim
\int _0^{\Lambda ^2}\frac{d\omega }{\sqrt{\omega ^2 + \mu ^2}}$,
where
$\mu $ is a dimensional parameter that sets the scale for the renormalized
theory. After incorporating this renormalization, equation (20) above in
(2+1) dimensions cam be written in a manifestly finite form.;
\begin{equation}
\int _0 ^{\infty} d\omega \frac{\tanh(\beta \sqrt{\omega ^2 + \Delta
^2}/2)
- 1}{(\omega ^2 + \Delta ^2)^{1/2}} = ln(\frac{\Delta
}{\mu }) = Lim_{\Lambda ^2 \rightarrow \infty }\left[\int _0^{\Lambda
^2}\frac{d\omega}{\sqrt{\omega ^2 + \mu ^2}} -
\int _0^{\Lambda ^2}\frac{d\omega }{\sqrt{\omega ^2 + \Delta ^2}}\right]
\end{equation}
Here $\omega = k^2$. In terms of the dimensional variables $y =
\frac{\Delta }{\mu}$  and $\zeta = \beta \Delta $, the renormalized gap
equation above becomes;
\begin{equation}
ln(y) = G(\zeta ) = 2\int _0^{\infty } d\theta \left[{1\over 1+e^{\zeta\cosh\theta}} \right]
\end{equation}
This is to be thought of as the equation of state for our system. The
other dimensionless quantities like the temperature in the units of the
binding energy $x = \beta \mu = \frac{y}{\zeta }$ can be recovered from
the parametric equation above. This equation is the same as the familiar
gap equation for the BCS superconductor, and the numerical values for the
physical quantities e.g. $kT_c/\Delta $ predicted by the above equation of
state are exactly equal to those of the usual BCS superconductor. 
Hence we observe, that in the absence of impurities, a dilute system of
electrons, interacting through arbitrarily small, short - ranged
attractive potentials,
will undergo a phase transition to a superconducting stage. 

Although the equations describing the phase transition are very
reminiscent of the BCS transition for metals, they do in fact describe a
different physical situation. This is evident if we note that the 
arguments that led to the mean field equation for this dilute system
of electrons are very special to the two dimensional nature
of the problem. Indeed  the nature of the divergence of the gap equation 
itself is different, in higher
dimensions, where the  equation does not admit any solutions for values
of the coupling constant less which are less than some critical value.

This is different from what happens in the theory of metals, at the
BCS transition point. As mentioned before, in the case of metals, one
deals with an effective theory of modes near the Fermi surface, which can
be reduced to a one dimensional theory with a linear dispersion relation which
leads to a similar logarithmic divergence. Moreover, this 
reduction is insensitive to the true dimensionality of the problem (the
exceptional case being D = 1) which makes the physics of the phase
transition blind to the true dimensionality of the metal being studied.         

It is clear from the discussion so far, that the model we are studying
now, does not admit of these simplifications, because of the
diluteness of the system. Since for our system,
$\frac{K_F}{\Lambda } \sim 0$, where $K_F $ is the Fermi momentum, and
$\Lambda $ is the typical energy scale of the problem, it is not
meaningful to look for a theory for the modes close to the Fermi surface.
This in turn ( as we pointed out above ) makes the phase
transition depend critically on the two dimensionality of the model, which
is an idealization. So we now probe the effects of a finite transverse
spatial direction on the model.
 
\section{The Analysis at Finite Thickness}
In this section we ask the question as to how thin the 
system has to be for it to be considered two dimensional.
 To keep matters simple, let us  investigate the situation at
zero temperature. In particular we want to see the transition from the
three dimensional case to the two dimensional one. Eqn (21) tells us that
in dimensions greater than two, the strong and weak coupling phases of the
system do not match smoothly, i.e. there is a critical coupling
$g^{-2}_cr \sim
\Lambda ^{D-2}$, below which the gap equation does not admit any
solutions. This is indicative of a phase separation between the strong and
weak coupling regimes. To understand the effects of a finite thickness
$L_3$ on the system, we will address the following questions.\\
a: What is the correct renormalization prescription to use for $g^-2$, for
small but non zero values of the the thickness $L_3$.\\
b:
Given
a finite thickness of the system $L_3$, and the fact that the system is in
the strongly coupled phase, how should $L_3$ approach zero, so that
the gap $(\Delta )$ remains finite.

At finite thickness, the non-relativistic dispersion relation reads
$\omega (k) = k^2 + (\frac{2\pi n}{L_3 })^2$, where n takes on integer  
values. So the gap equation at finite thickness is;  \begin{equation}
|g|^2\sum _n\int [d^2k][dk_0]\frac{1}{(k^2 + (\frac{2\pi n}{L_3})^2 )^2 +
\Delta ^2 + k_0^2)} = 1 \end{equation} $k_o $'s are the Matsubara 
frequencies which take on continuous values in $(-\infty ,
\infty )$ at $T = 0$.

Summation of the
series:\\ 

The series given by $S = \sum _n -f(i\omega _n)$, where
$f(i\omega n) = \frac{-1}{(k^2 - (i\omega _n )^2)^2 + k_0^2 + \Delta ^2} $
and $ \omega _n = (\frac{2\pi n}{L_3})$, can be evaluated by considering
the following contour integral;  \begin{equation} I = \oint\frac{dz}{2\pi
i}f(z)n(z), n(z) = \frac{1}{e^{L_3z} - 1} \end{equation} If the contour of
integration is taken to be the infinite circle centered at the origin,
then the integral vanishes, i.e. $\sum Res(f(z)n(z))  = 0$.  The poles of
the integrand are at $z = i\omega _n$, and at the four roots of $((k^2 -
z^2)^2 + k_0^2 + \Delta ^2 ))$.Let us denote the residues at the last four
poles by $R_i$. The residues at the poles labelled by the integer n are
$\frac{1}{L_3}f(i\omega _n)$. Hence $S = L_3 \sum _1 ^4 R_i$.  Summing the
series by computing the four residues, the gap equation at finite
thickness reads as;  \begin{equation} g^{-2} = \int
[d^2k][dk_0]\frac{L_3}{4\pi i b}\left[
\frac{\coth[(a-ib)^{1/2}L_3/2]}{(a-ib)^{1/2}} - \frac{\coth
[(a+ib)^{1/2}L_3/2]}{(a+ib)^{1/2}}\right]
\end{equation}
Here $a = k^2$ and $b = (k_0^2 + \Delta ^2 )^{1/2}$.

The
question now is to chose a suitable renormalization prescription for the
coupling constant. To do this we must first understand the nature of the 
divergence in the equation. As $\coth(x) \sim 1$, for $x >> 1$, equation
(26)
tells us that the nature of the divergence is linear for any finite value
of $L_3$ and logarithmic for $L_3 =0$. Hence at a finite thickness,     
choosing $g^{-2} \sim \int_{\Lambda } [d^2k]\frac{1}{(k^2 + \mu
^2)^{1/2}}$, i.e trading
the
dimensionless coupling constant for the dimensional parameter $\mu $
should make the problem manifestly finite. \\

To carry out the renormalization, we shall isolate the divergent part of
the integral appearing in the gap equation (27). Recalling that the
divergence in the integral remains the same when we let $L_3  \rightarrow
\infty $, i.e. let the hyperbolic cotangent go to $1$, we can add and
subtract, this divergent amount from the integral and rewrite it as;

\begin{eqnarray}
g^{-2} &=& \int \frac{[d^2 k ][dk_0]L_3}{4\pi i b}\left[
\frac{\coth[(a-ib)^{1/2}L_3/2]}{(a-ib)^{1/2}} - \frac{1}{(a-ib)^{1/2}} -
((a-ib) \rightarrow (a+ib)) \right] +\nonumber\\
& &+\int \frac{[d^2 k][dk_0]L_3
}{4\pi i
b}\left[ \frac{1}{(a-ib)^{1/2}} - ((a-ib) \rightarrow (a+ib)) \right]
\end{eqnarray}
Carrying out the integration over momenta (which are now finite), we have;
\begin{equation}
g^{-2} = \int \frac{[dk_0]}{2 i b }Arg\left[ \frac{1 -
e^{-(ib)^{1/2}L_3}}{1 - e^{-(-ib)^{1/2}L_3}}\right]  + L_3\int
\frac{[dk_0]}{b^{1/2}}
\end{equation}
The first integral in the equation above is finite while the second one
encodes the linear divergence of the problem. It diverges as $\sqrt{k_0}
$, and since $k_0 $ has the dimensions of $k^2$, it's divergence   
is
linear in the momentum cutoff. So the appropriate renormalization
prescription to use in this problem would be to let $g^{-2} =
L_3\int_{0}^{\Lambda ^2 }\frac{dk_0}{(k_0 ^2 + \mu ^2 )^{1/4}}$.
So the renormalized gap equation now is.
\begin{equation}
L_3'[\sqrt{\frac{\mu}{\Delta }}f(\Lambda ^2/\mu) - f(\Lambda ^2/\Delta )]
= G(L_3', \Lambda ^2 /\Delta)
\end{equation}
Here $f(a) = \int_0 ^a \frac{dx}{(x^2 + 1 )^{1/4}}$, $L_3' = \sqrt{\Delta
} 
L_3$, and $G(x, \Lambda ^2/\Delta) = \int_0^{\Lambda ^2/\Delta      
}\frac{dk_0}{\sqrt{k_0^2 + 1}}Arg\left[\frac{1 - e^{-\sqrt{i(k_0^2 +
1)^{1/2}}x}}{1 - e^{-\sqrt{-i(k_0^2 + 1)^{1/2}x}}}\right]$. The above
equation tells us that:
\begin{equation}
\sqrt{\frac{\Delta }{\mu }} =
\frac{1}{\frac{G(L_3',\Lambda ^2/\Delta)}{L_3'f(\Lambda ^2/\mu
)} + \frac{f(\Lambda ^2/\Delta )}{f(\Lambda ^2/\mu)}}
\end{equation}
In the limit $\Lambda \rightarrow \infty $, G remains finite, and the     
ratio of the two linearly divergent integrals approaches unity, and we are
left with a result that is insensitive to the finite size, i.e. $\Delta =
\mu $. So we see that for any finite value of $L_3$, the behavior of the 
system is the same as that of a truly three dimensional system in the
strongly coupled phase. This is indeed signalled by the renormalization  
prescription used above i.e $g^{-2} \sim \Lambda $, which is precisely of the order of the critical coupling in the full three
dimensional theory.

The interesting dependence on the finite size is in
the opposite direction, i.e. in the approach towards two dimensionality.  
So let us now ask the question, how should $L_3 $ approach zero 
(when the cutoff is removed) so that   
 $\sqrt{\frac{\Delta }{\mu }} $ remains finite ? To answer this question,
we revert back to Equation (26), and change the renormalization
prescription for $g$. We let $g$ be renormalized in the way it would be for
a truly two dimensional theory, i.e. 
$g^{-2} = \frac{1}{4\pi}\ln(\frac{2\Lambda ^2}{\mu}) $, and estimate the
 allowed range of values for $L_3$, which is now
thought of as a function of the cut-off. Expanding the hyperbolic
cotangents in  equation (26) to first order in $L_3$, we get,
\begin{equation}
\frac{1}{4\pi}\ln(\frac{2\Lambda ^2}{\mu}) = \int _{\Lambda
}[d^2k][dk_0]\frac{1}{k^4 +
k_0^2 + \Delta ^2} + \frac{1}{3}L_3^2\int _{\Lambda }[d^2k][dk_0]
\frac{1}{\sqrt{k_0^2 + \Delta
^2}}
\end{equation}
or, in terms of dimensionless quantities,
\begin{equation}
\ln(\frac{\Delta }{\mu }) = C L_3^2\Lambda ^2\int_0^{\Lambda
^2/\Delta}\frac{dx}{\sqrt{x^2 + 1}}
\end{equation}
Here C denotes a positive number (whose precise value is unimportant).
Hence from (31) above, we have,
\begin{equation}
\Lambda L_3 = \frac{\ln(\frac{\Delta }{\mu })}{\sqrt{C\ln(\frac{2\Lambda
^2}{\Delta })}}
\end{equation}
The equation above has the expected behavior, i.e, $L_3 \rightarrow 0$,
as $\Lambda \rightarrow \infty$.
$\Lambda ^{-1}$ can be though of as the measure of the lattice spacing in
two dimensions. Hence we see that the theory will behave like a two
dimensional theory even for a finite thickness, as long as the transverse
thickness, $L_3$ measured in units of the lattice spacing in the two
dimensions is small. An estimate of this smallness is provided in the
equation above.

\section{Justification of the mean field theory, a renormalization group
analysis}

In the previous sections we carried out the mean field analysis of a
system of non - relativistic Fermions interacting through a short ranged
attractive potential, by constructing the corresponding Landau - Ginsburg
theory. At the level of the mean field theory, we identified a logarithmic
divergence in the problem, which was removed by a renormalization of the
coupling constant. The mean field analysis also predicted the existence of
a gap in the spectrum, for arbitrarily small values of the attractive
coupling. So the natural question that arises is that how can one justify
the validity of the mean field approximation, which leads to these non -
trivial consequences? In other words, we know of several examples where
the predictions of mean field theory are qualitatively wrong; the $D =
1$, Heisenberg chain at half filling\cite{Shankarbcs} is a good example of
this; examples
such as these provide us with a warning that motivate a surer
justification of the mean field analysis. In this section here, we shall
carry out a one loop renormalization group analysis of the theory, and
show that the BCS instability predicted by the mean field analysis is also
implied  by the renormalization group flow equations , and this lends
substance to
the qualitative nature of the predictions of the mean field theory.

The logistics for carrying out the renormalization group analysis for
our model are going to be as follows; we shall start with a two
dimensional, free, scale invariant theory, which we will
define with a fundamental cut-off $\Lambda $. The  Gaussian action is;

\begin{equation}
S_0 = \int[dk_0] \int [d\theta ]\int_0^{\Lambda }k[dk][\Psi ^{\dagger
}(k_0k)(ik_0 - k^2)\Psi (k_0k)]
\end{equation}

The  action is invariant under the scaling transformation $k
\rightarrow sk, k_0 \rightarrow s^2k_0, \Psi \rightarrow s^{-3} \Psi$.

Next we shall  introduce various scale invariant perturbations around the
free
action, and study the evolutions of these perturbations under the RG  
transformations. To generate the RG transformations, we shall split up the
phase space (the k space ) into `high' and ` low' degrees of freedom, the
high and low degrees of freedom being defined as, $\Psi _{high} = [\Psi
(k)$, for $(\Lambda /s \leq k \leq \Lambda )$ and $0$ otherwise$]$, $\Psi
_{low} = [\Psi (k)$,
for $(0 \leq k\leq \Lambda /s )$ and $0$ otherwise $]$. $s$ is a real 
positive number greater
than one.
Following this we'll proceed to integrate out the high degrees of freedom,
and get an effective theory for the low modes. This process will generate
the RG flow equations for the various couplings that perturb the free
action. Of special interest to us is the flow corresponding to the four  
point interaction, as that encodes the possibility or (lack of) a BCS type
of instability in the system.

Before we proceed to carry out the renormalization group transformation
for the four point function, it is worth pointing out that unlike the
usual BCS theory, we are not building an effective theory for the modes
close to the Fermi surface, indeed we are interested in the scenario,
where the system is sufficiently dilute such that $\frac{K_F}{\Lambda
}<<1$. In the case of the BCS theory, since the physically
interesting degrees of freedom are the modes close to the Fermi surface, 
one is justified in linearizing the energy momentum dispersion relation   
around the Fermi energy and  keeping the contribution only from the
direction
normal to the Fermi surface, as these correspond the degrees of freedom of
significance to the RG transformations. The degeneracy of the angular
directions, and the linearization of the dispersion relation allows one
to reduce the theory to a $1+1$ dimensional  theory with a relativistic
dispersion relation, involving
N species of Fermions, where N is a large number proportional to the area
of the Fermi surface measured in units of the cutoff.
These generic arguments carry through in any dimension greater than or   
equal to two, which is an explanation for the fact that all metallic
systems, irrespective of the number of space dimensions face the BCS
instability in the limit of $T \rightarrow 0$.

In the case we are interested in such simplifications are absent, although
the BCS instability is present, however in  a way, which is special to the
two dimensional nature of the problem.

To see this, let us start with a generic four point coupling;
\begin{equation}
S_{int} =\int\Pi _i [d\vec{k_i}][dk_{0i}]U(\vec{k_i})\Psi^{\dagger  
}(4)\Psi^{\dagger}(3)\Psi (2)\Psi (1)\delta (\vec{k})\delta (k_0)   
\end{equation}

In the above interaction the number 1...4 denote the momenta and
Matsubara frequencies. U is the four point coupling function, which in
general will have a dependence on the momenta. The delta functions ensure 
than the momenta and the Matsubara frequencies corresponding to 1 and 2
equal those of 3 and 4. All the momentum integrals run between 0 and
$\Lambda $. If we expand the four point function in a Taylor series, only
the constant part ($U_0$) turns out to be of importance, as the terms with
dependence on the momenta, correspond to irrelevant couplings, i.e, only
the constant part leads to a four point interaction invariant under the
scaling defined above.

If we now split up the fields into the high and low modes, then the one
loop corrections to the coupling constant are given by the three diagrams
above (fig 1), where the internal loop momenta correspond to the high
degrees of
freedom, i.e $\Lambda /s \leq$ loop momenta $\leq \Lambda $.  i.e.

\begin{figure}
\centerline{\epsfxsize=6.truecm\epsfbox{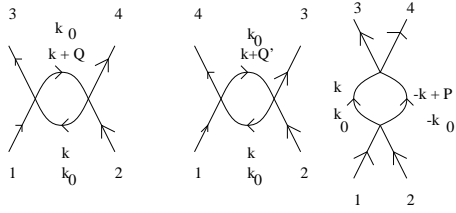}}
\caption{ Graphs Contributing to one loop $\beta $ function}
\end{figure}  

\begin{eqnarray}
\delta U_0 &=& u_0^2\int _{-\infty }^{\infty }dk_0 \int _{\Lambda
/s}^{\Lambda }
[d^2k]\int d\theta  [\frac{1}{(ik_0 - k^2)(ik_0 - (k+Q)^2)} +
\frac{1}{(ik_0
- k^2
)(ik_0 - (k+Q')^2)} - \nonumber \\\
& &\frac{1}{2}\frac{1}{(ik_0 - k^2)(-ik_0
-(P-k)^2)}]
\end{eqnarray}

It is now obvious that the first two diagrams are identically equal to
zero as the $k_0$ integrals vanish for these diagrams. Both the poles are
on the same  half of the complex $k_0 $ plane because $(k + Q)^2$  and $(k
+ Q')^2$ being quadratic functions are always positive for all values of
momentum transfer. Hence we can always close the contour in the other half
plane and this makes these integrals vanish. The third integral however is
non zero, and produces a flow, which is given by;
\begin{equation}
\beta = \frac{dU_0}{dt} = \frac{-U_0^2}{4\pi}
\end{equation}
Here $t = -\ln s$, and the solution to the flow equation is,
\begin{equation}
U_0(t) = \frac{U_0(O)}{1 + \frac{t}{4\pi}U_0(0)}
\end{equation}

Thus we find that for an attractive microscopic coupling, (corresponding
to $U_0(0) < 0$), a BCS instability is inevitable. Hence the divergence
of
the coupling constant that showed up in the mean field analysis is not an
artifact of the mean field approximation, but is indeed a true signal of
the onset of a BCS transition.

We would like to emphasize the difference of this coupling constant
renormalization with its counterpart in the theory of metals. In metallic
systems, the constraint that all momenta lie very close to
the Fermi
surface imposes a stringent dependence of the coupling constant on the  
transverse momenta, i.e on the directions orthogonal to the radial
directions in momentum space. Moreover the only couplings that contribute
to the flow are the ones for which $\vec{k_1} = -\vec{k_2}$, and
$\vec{k_3 } = - \vec{k_4 }$, and  $U_0 = U_0(\vec{k_1}.\vec{k_3})$, i.e,
it
is a function of an angle. When this function is decomposed into it's
various angular momentum components, the contribution from the first two 
diagrams (which are non zero, as the dispersion relation is linear in the
momentum) couple the flows corresponding to the various angular momentum
sectors  (the Kohn - Luttinger effect\cite{Luttinger1, Luttinger2,
Baranovetal}), in a manner that drive (for certain values of the angular
momentum) couplings which might have been repulsive to negative values.  
These couplings then  lead to the BCS instability according  to the
flow equation
(37)
given above.

Hence it is clear that the system we are presently interested in is
different from the usual BCS superconductor, although it also possesses a
similar
ground state. But there is no analogue of the Kohn Luttinger effect for
this
system, however a coupling that is attractive to start with will be driven
by the RG flow to produce a BCS like ground state. Moreover, the
arguments and the analysis we given above are very special to the two
dimensional case as was shown in the previous section.

\section{Conclusion:}
We conclude that a dilute, two dimensional system of electrons, 
interacting
through a short ranged attractive potential will undergo a  phase
transition to a superconducting regime. This phase transition is different
from the usual BCS transition in several respects.

a: The phenomena
described here is very special to the  two
dimensionality of the problem. 

b: The  phase transition  is from a
semiconducting /
insulating phase to a superconducting regime. 

c: The diluteness of the system plays a very significant role in the
analysis, as that is what warrants the use of the quadratic dispersion
relation, for which the critical dimension is $d = 2$. 

On a speculative note, we would like to emphasize that these phenomena may
be of relevance towards understanding the behavior of high $T_c$
superconductors\cite{Abrahamsetal:metal2d, Phillips1}. 

In
dimensions greater than two we find that this
kind superconductivity will not arise for arbitrarily small values of the
electron -
electron coupling, and that the coupling will have to be of greater than
some critical value $g_{cr}$
for the phase transition to take place. If one allows for a finite
thickness of the material being studied, then we find that there are
stringent bounds on the allowed range of thickness if the phenomena
characteristic of two dimensions is to survive; in fact we find that the
square of the thickness can exceed the lattice spacing
in two
dimensions only by logarithmic amounts.   

{\bf Acknowledgement:} We thank G.Krishnaswami and Y.Shapir for useful
discussions. This work was supported in part by the US Department of
Energy, Grant No. DE-FG02-91ER40685

\bibliography{abhishekbib}

\end{document}